# What is prompt literacy? An exploratory study of language learners' development of new literacy skill using generative AI

Yohan Hwang[a], Jang Ho Lee[b]*, Dongkwang Shin[c]*

[a]Department of English Language and Literature, Jeonju University, Jeonju, Republic of Korea; [b]Department of English Education, Chung-Ang University, Seoul, Republic of Korea; [c]Department of English Education, Gwangju National University of Education, Gwangju, Republic of Korea

*Corresponding Author

## ABSTRACT

In the current study, we propose that, in the era of generative AI, there is now a new form of literacy called "prompt literacy," which refers to the ability to generate precise prompts as input for AI systems, interpret the outputs, and iteratively refine prompts to achieve desired results. To explore the emergence and development of this literacy skill, the current study examined 30 EFL students' engagement in an AI-powered image creation project, through which they created artworks representing the socio-cultural meanings of English words by iteratively drafting and refining prompts in generative AI tools. By examining AI-generated images and the participants' drafting and revision of their prompts, this study demonstrated the emergence of learners' prompt literacy skills. The survey data further showed the participants' perceived improvement in their vocabulary learning strategies as a result of engaging in the target AI-powered project. In addition, the participants' post-project reflection revealed three benefits of developing prompt literacy: enjoyment from manifesting imagined outcomes; recognition of its importance for communication, problem-solving and career development; and the enhanced understanding of the collaborative nature of human-AI interaction. These findings suggest that prompt literacy is an increasingly crucial literacy for the AI era.

**Key words:** AI literacy, generative AI, prompt literacy, ChatGPT, vocabulary learning

## 1. Introduction

With technological advancement, the way knowledge is acquired and shared is being transformed. Consequently, the literacy skills required of students are evolving in tandem with technological breakthroughs (Jewitt, 2012). In traditional literacy learning environments, students primarily engage in learning through reading and writing activities. However, with the emergence of the digital technology era (Gilster, 1997), digital literacy has expanded how learning is reinforced through various visual elements, such as images and videos (Debes, 1969). In a similar vein, the significance of multiliteracies, where students analyze and interpret multimodal learning content, has been emphasized in the academic literature (The New London Group, 2000). Moreover, recent immersive digital technologies, such as augmented reality (AR) and virtual reality (VR), have enabled experiential knowledge acquisition. This allows learners to experience literary content in simulated environments (Lee, 2023; Wu et al., 2023). Consequently, this shift has led to the reconceptualization and expansion of experiential educational theories, including Dewey's experiential learning theory (1938) and Dale's cone of learning model (1969), within the context of virtually enhanced reality (Hwang et al., 2023).

As AI technology has progressed, the era of AI literacy has begun (see Ng et al., 2021 for review). In this context, "literacy" refers to the understanding and capability to interact with, utilize, and critically evaluate AI systems and their implications. This concept goes beyond just knowing how AI works technically; it encompasses AI's broader societal, ethical, and practical aspects. Recently, as we have begun moving from the age of descriptive AI to that of generative AI, there has been a rise in the development of emergent literacy skills (Bozkurt, 2023; Su et al., 2023). Given that generative AI results can vary based on text-based input, namely the "prompt," research is urgently needed on how students can effectively draft and modify prompts, and how they can evaluate the resulting output to improve their literacy skills.

To address this research gap, the present study introduces the notion "prompt literacy," which refers to the ability and skill to generate precise input for generative AI, interpret the output, and modify prompts to achieve desired results (Hwang, 2023). Building on this new concept, we aim to explore how prompt literacy can be conceptualized and operationalized in the context of second language (L2) learning, in which English as a Foreign Language (EFL) students engage in creating visual artwork representing socio-cultural and historical origins of English vocabulary by interacting with generative AI.

## 2. Background

### 2.1 Conceptualizing prompt literacy alongside technological development

The development of emergent literacy skills is closely linked to the sociocultural landscape and technological development of the era. As Jewitt (2008) highlights, "there is a need to approach literacy practices as an inter-textual web of contexts and technology, rather than isolated sets of skills and competences" (p. 47). Traditional literacy, in its conventional sense, is typically related to reading and writing skills, with its application primarily confined to paper and books (Hull & Moje, 2012). However, literacy now serves as a means for communicative interaction with other texts, events, and people, with meaning being co-constructed through individual–context interaction (Johns, 1997). In other words, the development of computers and the internet has shifted the environment in which people consume text, namely from offline to online; as such, the conventional concept of literacy has expanded into a new form known as digital literacy (Gilster, 1997). This transformation highlights the changing landscape of how people engage with texts and underscores the need for a broader set of literacy skills to navigate the digital world.

As we have already entered the era of digital literacy, the importance of skills now extends beyond textual comprehension and manipulation to encompass the ability to understand and utilize various forms of media (e.g., images and videos) as visual literacy (Debes, 1969). Additionally, there is an increased emphasis on the capacity to gather, process, infer, and judge information found on the internet as information literacy (Bawden, 2001). In other words, literacy is no longer confined only to text processing; the emphasis now lies in the ability to engage with media-rich digital information, effectively use it, and create relevant content (Andretta, 2007; Burniske, 2007).

Furthermore, with AI technology advancing at an unprecedented rate, the scope of digital literacy has expanded to AI literacy (Ng et al., 2021). It encompasses the knowledge, skills, and attitudes necessary for engaging with AI across various aspects of life (Kong et al., 2021; Steinbauer et al., 2021). During the expansion period from digital to AI, the primary emphasis was placed on understanding the principles and operation of AI technology, resulting in a surge of interest that emphasized computational thinking skills and led to a coding frenzy in the education sector (Liang et al., 2020). However, with the recent emergence of generative AI models capable of autonomously generating code, learners can accomplish a wide range of tasks using low- and no-code approaches. In this circumstance, much like the evolution of literacy in the past in response to other technological advancements, the evolution of technology has necessitated a further progression in literacy skills beyond AI literacy in tandem with AI's new capabilities.

In this study, we propose that this skill is referred to as "prompt literacy." The process is described in the following way: a prompt encompasses text or information input into generative AI models. Then, through prompts, AI comprehends the provided variables and generates appropriate outputs. The more detailed and specific the prompt, the more accurate the AI-generated results. Importantly, prompt literacy goes beyond just providing initial input; it also involves evaluating the accuracy of the results produced by generative AI and continually modifying the prompt until the desired information is obtained (Hwang, 2023). Furthermore, it includes all actions taken to create the desired output by combining various forms of information provided by different generative AI programs. Figure 1 illustrates the status of prompt literacy, in relation to other relevant concepts of literacy.

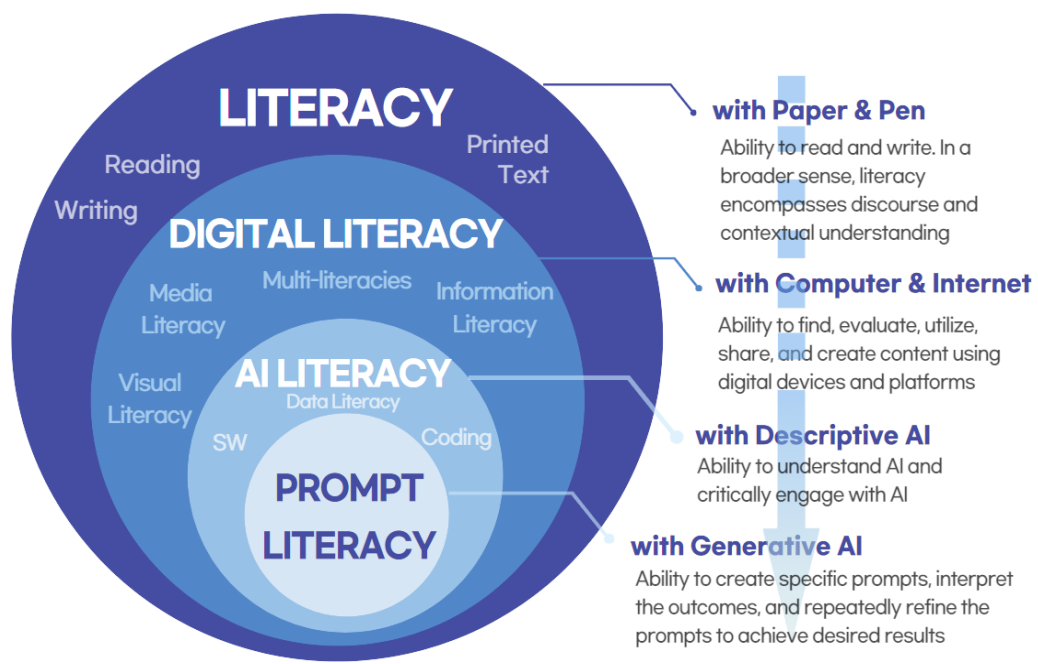

**Figure 1.** The concept of prompt literacy

Recently, there has been a growing interest in research aimed at advancing information literacy through the practice of prompt engineering. In fact, various industries actively seek individuals with expertise in prompt engineering, and researchers are exploring innovative approaches to teach this skill to students (Giray, 2023; Oppenlaender, 2023). In this context, Lo (2023) has introduced a novel framework designed to serve as a guiding tool for prompt engineering. This approach incorporates the principles of CLEAR (conciseness, logic, explicitness, adaptability, and reflectiveness) to enhance prompt efficacy. While this study emphasizes the importance of crafting accurate and sophisticated prompts, it does not address how learners interpret the multimodal results created by generative AI or how to revise their prompts accordingly. In summary, prompt literacy differs from prompt engineering in that it puts more emphasis on how to interpret generative-AI outcomes, how to set effective prompt-making strategies for better outcomes, and how to refine prompts until the intended results are generated.

## 2.2 Operationalizing prompt literacy in cognitive vocabulary learning approaches

According to Dale's (1969) cone of learning model, students remember differently depending on how they learn (i.e., reading, hearing words, viewing images, watching videos, looking at an exhibition, watching a demonstration, participating in hands-on workshops, designing collaborative lessons, simulating a model, performing an experiment). Among other avenues, learning through linguistic symbols and gaining experiential knowledge is more effective for cognitive development than just reading and writing down information (Bruner, 1983). When learning English vocabulary, it is important for students to understand the words' meaning through iconic and symbolic representation. In this vein, one of the most frequently-cited and widely-practiced approaches to introducing and teaching new vocabulary to L2 learners is to present it along with visual aids that have some connection to the target vocabulary (Gersten & Baker, 2000; Rousseau et al., 1993). In the L2 domain, the effectiveness of using visual supports, particularly pictures, has been examined largely by reading research on multimedia glossing (e.g., Al-Seghayer, 2001; Kost et al., 1999; Yanguas, 2009). The results of these studies have generally supported using pictures for incidental L2 vocabulary acquisition, thereby confirming its pedagogical value.

Presenting visual aids for L2 vocabulary can be supported by two theories of human cognition: dual coding theory (Paivio, 1986, 2010) and Mayer's (2009) cognitive theory of multimedia learning (CMTL). The former suggests that humans have two separate but interconnected systems for processing and representing information: verbal and visual. This theory further predicts that human cognition is likely to process and retrieve target information better when it is presented in both verbal and visual formats. In contrast, the latter emphasizes the importance of including multimedia components (e.g., pictures) in instructional materials, a practice that can enhance learning by engaging learners in different cognitive processes. Although dual coding theory focuses more on the cognitive process of information and CMTL pays more attention to building effective instructional materials, both theories point to the importance of presenting information in both verbal and visual channels for effective processing and acquisition.

Despite the aforementioned theoretical support for visual aids in L2 vocabulary learning,

such an approach may have some limitations. According to Nation (2001), one of the limitations may lie in the nature of pictures themselves, namely that they generally contain a lot of information. Because of this, Nation suggests that "it may be necessary to present several examples so that learners can determine the essential features of the concept or accompany … pictures with focusing information (p. 85). Indeed, when it comes to teaching the sociocultural (rather than surficial) meaning of some target words (or expressions), a careful selection of images is deemed necessary. In addition, teacher-provided materials may not enable learners successfully to encode and retrieve information related to the target vocabulary, especially if those images are beyond the learner's cognitive scope (see Cohen, 1987; Paivio & Desrochers, 1981 for a similar view). It may be the case that learner-generated images may better facilitate their vocabulary learning, although having learners produce their own images for target vocabulary might be less effective if they have limited drawing abilities or lack digital literacy skills. This could especially be true when trying to represent visually words' sociocultural meaning. These limitations, however, can be compensated for by employing a text-to-image generator, with which learners and teachers alike can instantly create images representing vocabulary by drafting and revising prompts according to educational purposes.

Bruner (1983) emphasized the importance of the mode of representation in cognitive knowledge acquisition, simplifying Dale's (1969) cone of learning model into three stages: enactive, iconic, and symbolic representation. When learning how to represent visually L2 vocabulary, operationalizing prompt literacy can help cognitive development across these stages. Just as a young child acquires experiential knowledge through the three modes of representation, today's students in the AI era can acquire new knowledge through prompt literacy.

First, in the enactive stage, students can create prompts and then input them into a generative AI program to search for a word's meaning and usages. For instance, using ChatGPT, students can explore a word's historical origins or sociocultural meaning and navigate its actual usage in sentences, discourses, and contexts. This stage also includes a series of activities in which students modify their prompts to achieve desired results. This process closely resembles the enactive approach, as students intuitively grasp the meaning through their prompts' linguistic action. As Bruner (1983) highlighted, this enactive representation aligns with the manipulative representation of past events. Students proficient in prompt literacy can experience different outcomes based on the iterations of their previous actions when creating and refining prompts. Here, "manipulation" also refers to skillfully handling prompts to guide the generative AI's actions.

Second, the iconic representation stage involves creating internal visual imagery to represent concretely a word's implied meaning. In a traditional learning environment, even if students cannot physically see the actual object, they should be able to visualize it vividly in their minds. However, as mentioned earlier, this approach has the drawback of limiting students' imaginations in connecting to tangible outcomes. In this context, generative AI can assist students in generating visual aids, images, or diagrams for abstract concepts. In other words, when learning L2 vocabulary through prompt literacy, tools such as text-to-image or text-to-video generators can produce iconic representations that facilitate students' cognitive development.

Lastly, according to Bruner's (1983) original theory, symbolic representation is a stage that delves into the interplay between language and images. In conjunction with prompt literacy, students can go beyond merely understanding the meanings of words or images themselves. They can also gain an abstract understanding of how different prompts lead to various outputs and how they interact. This process allows students to formalize the structure and templates of prompts as a symbolic system through creating and refining prompts necessary to achieve their desired results.

To offer empirical evidence for operationalizing prompt literacy within an educational context, this study explores the experiences and perceptions of students who have utilized generative AI tools, including ChatGPT and Bing Image Creator, to create visual artwork. Given the theoretical advantages of presenting visual aids in L2 vocabulary instruction, as well as the aforementioned potential of prompt literacy, students were asked to engage in image generation on their own by interacting with generative AI to produce the most appropriate image that learners could use to associate with the target vocabulary. They did this by drafting and revising the prompts iteratively. The following is the primary guiding research question driving the current study: How do prompt literacy skills develop when EFL students engage in an AI-powered vocabulary-image creation project, and would such a development impact their subsequent vocabulary learning and engagement with generative AI?

## 3. Method

### 3.1 Participants and research context

This study documents the experiences of 30 college students who engaged in a generative AI-based image creation. They majored in English Language and Literature at a private university, in Jeonju, South Korea, and participated in a course titled "Digital Literacy and Humanities" during the first semester of the 2023 academic year. The course adopted an interdisciplinary approach, exploring the intersection of generative AI and the Humanities, and it underscored the progressive interaction between humans and AI. Specifically, students gained insight into the evolution of prompt literacy, tracing its development from traditional literacy through digital literacy to AI literacy. Furthermore, they probed the selection of appropriate programs for proficiently employing prompt literacy; this included the precise input of prompts, the interpretation of results produced by generative AI, and the methods for modifying prompts. For the vocabulary-image creation project, in this course, the participants studied 120 English words included in a course book titled *Story of English Words* (Cho, 2020). This text discusses the words' socio-cultural influences and historical origins that led to their current meanings and uses. For example, *Story of English Words* narrates why "spam" transitioned from meaning "ham" to "junk mail" or how "bug" evolved from an "insect" to signify a "machine malfunction." After reading such explanations for all the words introduced in the book, the participants were asked to choose three words each and to visualize them by employing generative AI.

The overall procedure was as follows. The participants first confirmed their understanding of the selected words' meanings by using ChatGPT (https://chat.openai.com/). Then, they conducted research to collect additional information and summarized their findings. Next, they sketched an image either by hand or with digital tools. Afterward, they drafted a prompt for a generative-AI-produced image, inputting it into the Bing Image Creator (https://www.bing.com/create) to produce an image. Importantly, the prompt input and image generation process were not limited to a single attempt; participants revised the prompt until the AI-generated image closely matched their initial sketch and imagination. They then shared the images' iterations, which changed based on each prompt revision.

For instance, one participant selected the word "siren." This student first drew a ship in the sea accompanied by a mermaid to visualize this word. Then she crafted a prompt for the Bing Image Creator. The initially drafted prompt was: (1) Draw a European ship sailing in the heart of the sea, **with a mermaid close by**, singing to entice the ship's crew. However, the illustration showed the mermaid positioned directly beside the ship, which differed from her original idea. She then rephrased the prompt to have the mermaid singing from the shore, watching the distant ship: (2) Draw a European ship sailing in the heart of the sea, **with a mermaid on the shore**, singing to entice the ship's crew. Surprisingly, the new illustration featured an additional ship on the shore. To address this and further specify her vision, she adjusted the prompt one again to clarify the mermaid's location and provide more details about the ship and the sea: (3) Draw a **huge** European ship sailing in the heart of the **blue ocean with its crew atop the deck, while a mermaid serenades from rocks along the coastline** (see Figure 2 for the images generated with each prompt).

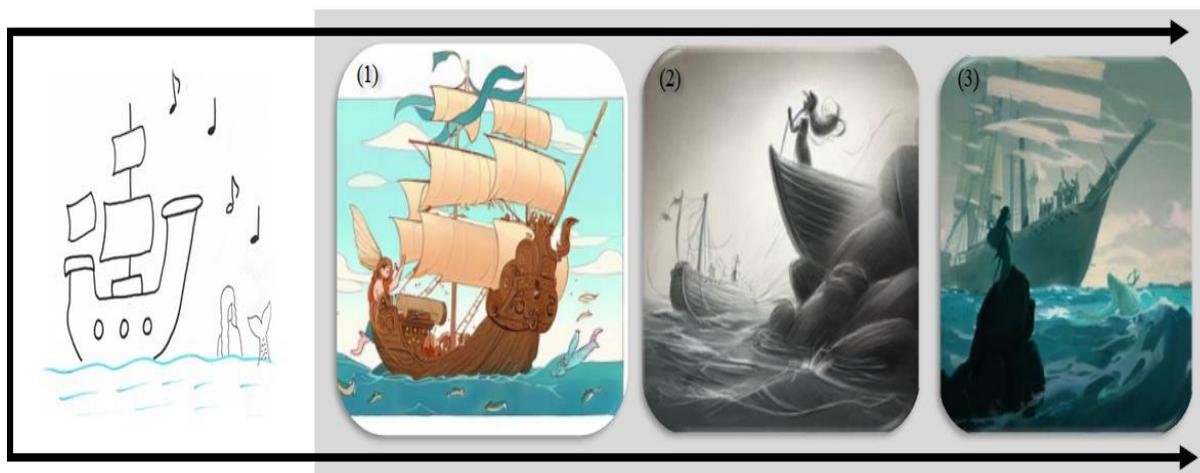

**Figure 2.** Sample images associated with "siren"

Another participant chose the word "text." She first learned about its origin from *Story of English Words*, discovering that the word "text" originates from the ancient Greek and Roman appreciation of the delicacy of women. She further probed information by talking to ChatGPT, learning that because making clothes was an essential skill for women during Roman times, fabric was called "textile." Teachers of that era encouraged male students to learn from the delicacy of women, leading to the word "textile" evolving into "text." She then sketched an image depicting the process of generating text as being similar to using a loom to create clothing made of words. The first prompt that she inputted into the Bing Image Creator was: (1) **Fabric** being woven using **a loom with letters** from all over the world woven into fabric. She was satisfied with the initial image, but to emphasize the depiction in the book where a woman uses a loom to create something delicate, she modified the prompt accordingly: (2)

**Woman using a loom** to weave fabric with letters from all over the world and then the **fabric turning into a book.** She then realized that the resulting image did not resemble a woman from ancient Greece and made adjustments to her prompt in terms of the woman's appearance: (3) **Woman in ancient Rome** using a loom to weave fabric with letters from all over the world and then the fabric turning into a book (see Figure 3 for the images generated with each prompt).

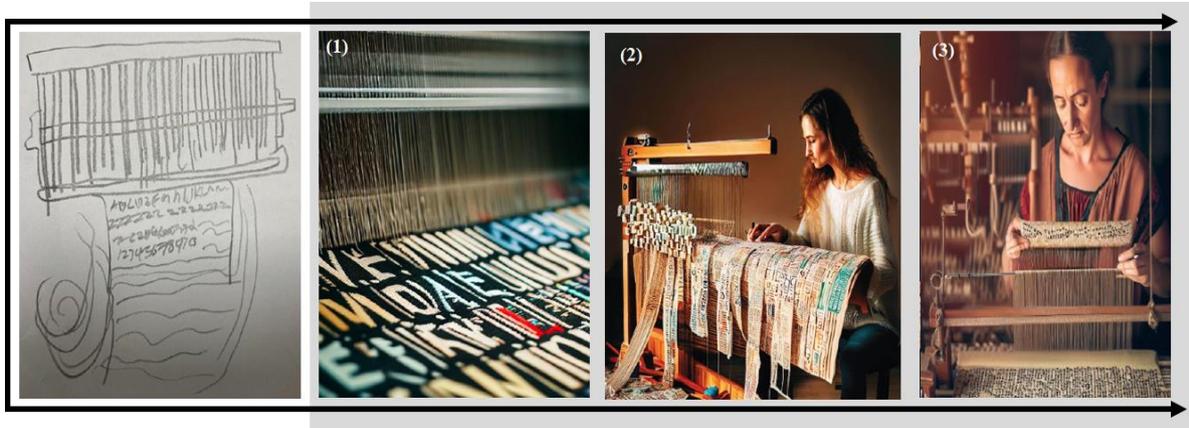

**Figure 3.** Sample images associated with "text"

In general, it was found that the image generation process for the participants was highly iterative, involving several revisions of the original prompts.

### 3.2 Data collection and analysis

The present study includes three types of data: (1) the artwork (images) generated by the participants; (2) survey questions using a 7-point Likert scale (see Table 1); these questions examined the participants' self-perceived development of cognitive vocabulary learning strategies via the AI-powered vocabulary-image creation project, adapted from Gu and Johnson's (1996) Vocabulary Learning Questionnaire (VLQ); (3) and the participants' written responses obtained from a reflection paper describing their experience using generative AI programs to create images, and their own definition of prompt literacy and its pedagogical value.

**Table 1.** Survey questions

| Category | Strategies | Constructs | Questions |
|---|---|---|---|
| Encoding | Visual encoding | Visualization | 1. After participating in the vocabulary-image creation project with generative AI, I visualize related images when learning new words. |
| | Use of word-structure | Hidden Meaning | 2. I think participating in the vocabulary-image creation project with generative AI helped me grasp the hidden meanings of English words. |
| | | Prefix & Suffix | 3. After participating in the vocabulary-image creation project with generative AI, I now consider prefixes, suffixes, and etymologies when learning new words. |

| | | Origin | 4. After participating in the vocabulary-image creation project with generative AI, I now intentionally think about the origin of the words. |
|---|---|---|---|
| | Contextual encoding | Use in context | 5. After participating in the vocabulary-image creation project with generative AI, I feel I would want to understand the context in which the word is used. |
| | | Use in sentence | 6. While preparing for the vocabulary-image creation project with generative AI, I became able to think of expressions or sentences containing the word. |
| | | Use in situation | 7. While preparing for the vocabulary-image creation project with generative AI, I became able to imagine real-life situations where the word is used. |
| Activation | Activation | Imagination | 8. After participating in the vocabulary-image creation project with generative AI, I now try to use newly learned words in imaginary situations in my mind. |

For data analysis, the artwork generated by the participants were closely examined in terms of how the learners conceptualized and operationalized their prompt literacy in the AI-powered vocabulary-image creation project. We initially analyzed how the results changed depending on the prompt modifications. Then, we thematized which strategies the participants had employed to achieve their desired images. Regarding the post-project survey, descriptive statistics were calculated to assess participants' self-perceived development of vocabulary learning strategies via the target project. For the data obtained from the participants' reflection papers, we employed a conventional approach to content analysis (Hsieh & Shannon, 2005), through which learners' reflections on the various aspects of the AI-powered image creation project and prompt literacy were condensed into common content categories through an iterative coding procedure (Saldaña, 2013).

## 4. Result

### 4.1 Developing prompt literacy to generate the intended results

Overall, two major strategies emerged in terms of how the participants drafted, modified and interpreted prompts to express the meanings of the English words they had learned. First, the participant mainly crafted and adjusted prompts to depict more precisely the context in which the target word is used according to their sociocultural and historical backgrounds. During this process, they employed various linguistic phrases to describe the background that could represent the word's meaning more accurately. As a result, they often modified verbs, adjectives, prepositions, and colors given in the earlier prompts. Second, the participants emphasized the actions of characters/objects that related to the words' histories. They focused on describing what the imagined characters are doing, holding, wearing, and more. When modifying the prompts, it was observed that the participants revised the use of verbs frequently to this end.

For example, one participant studied how the word "client" originated from "a person who bows down," indicating a subordinate in ancient Rome, but that has now changed to imply a high position as in the phrase "the customer is king." The learner then collaborated with ChatGPT and Bing Image Creator to generate an image that could illustrate this background (see Figure 4). Initially, she asked for an image in which many people are serving a client, bowing down to them as if they were a highly-revered leader. However, when the image depicted people turned away from the client, in the second attempt, she highlighted the action "people facing forward." In the second image, although the client was facing forward, it did not convey the exact sense that a client is in the higher position than the others in the background. Therefore, in the third attempt, she added the prepositional phrase "on the higher position" to generate an image that could show the client having a higher status.

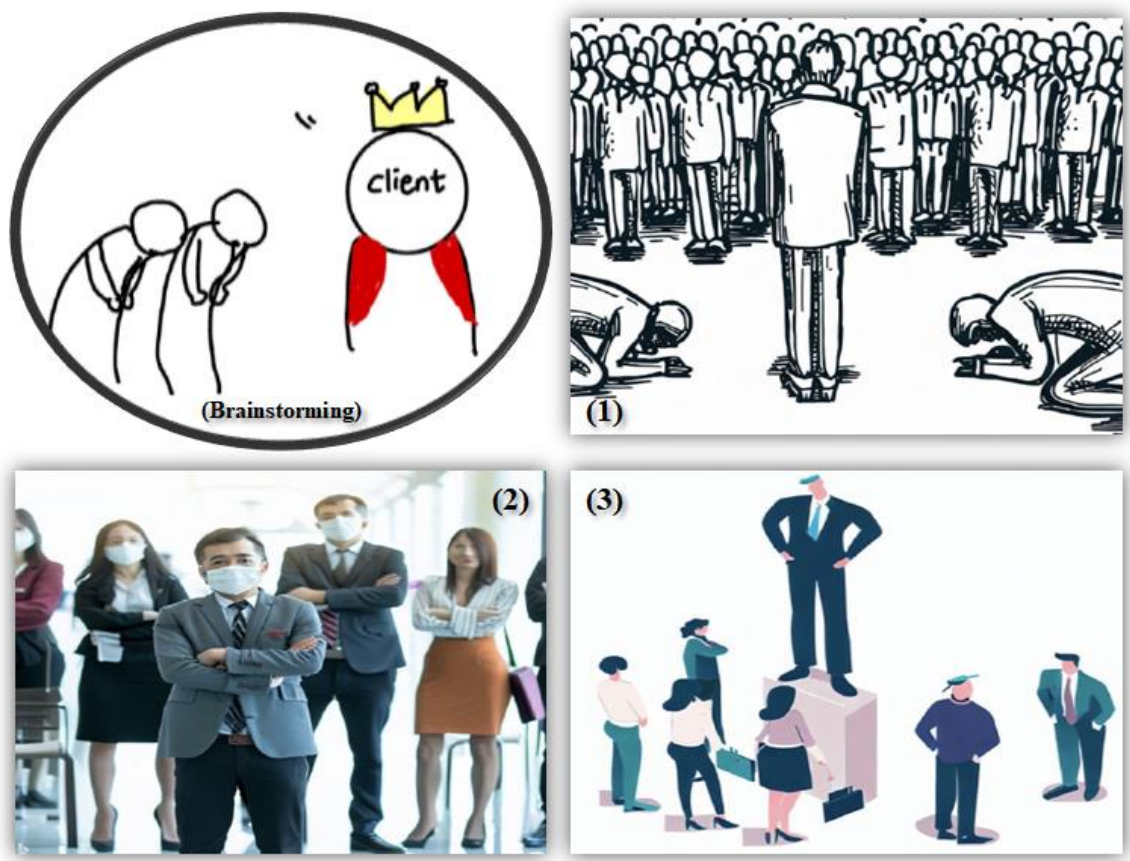

**Figure 4.** Sample images associated with "client"

Another participant chose the word "California." Through reading the course book and conversing with ChatGPT, she learned that the name of the American state of California was derived from Queen Calaphia, who ruled an unknown island filled with treasures and inhabited solely by female warriors. Then, she requested a prompt of this mysterious, treasure-filled island where the queen lived. However, when the initially generated image did not depict the actual inhabitants of the island accurately, she modified the prompt to emphasize the female warriors. The following image portrayed better the warriors inside the island, but she realized that the background did not capture California's bright essence. The prompt was then adjusted to highlight the lush vegetation in the background, leading to the final image (see Figure 5).

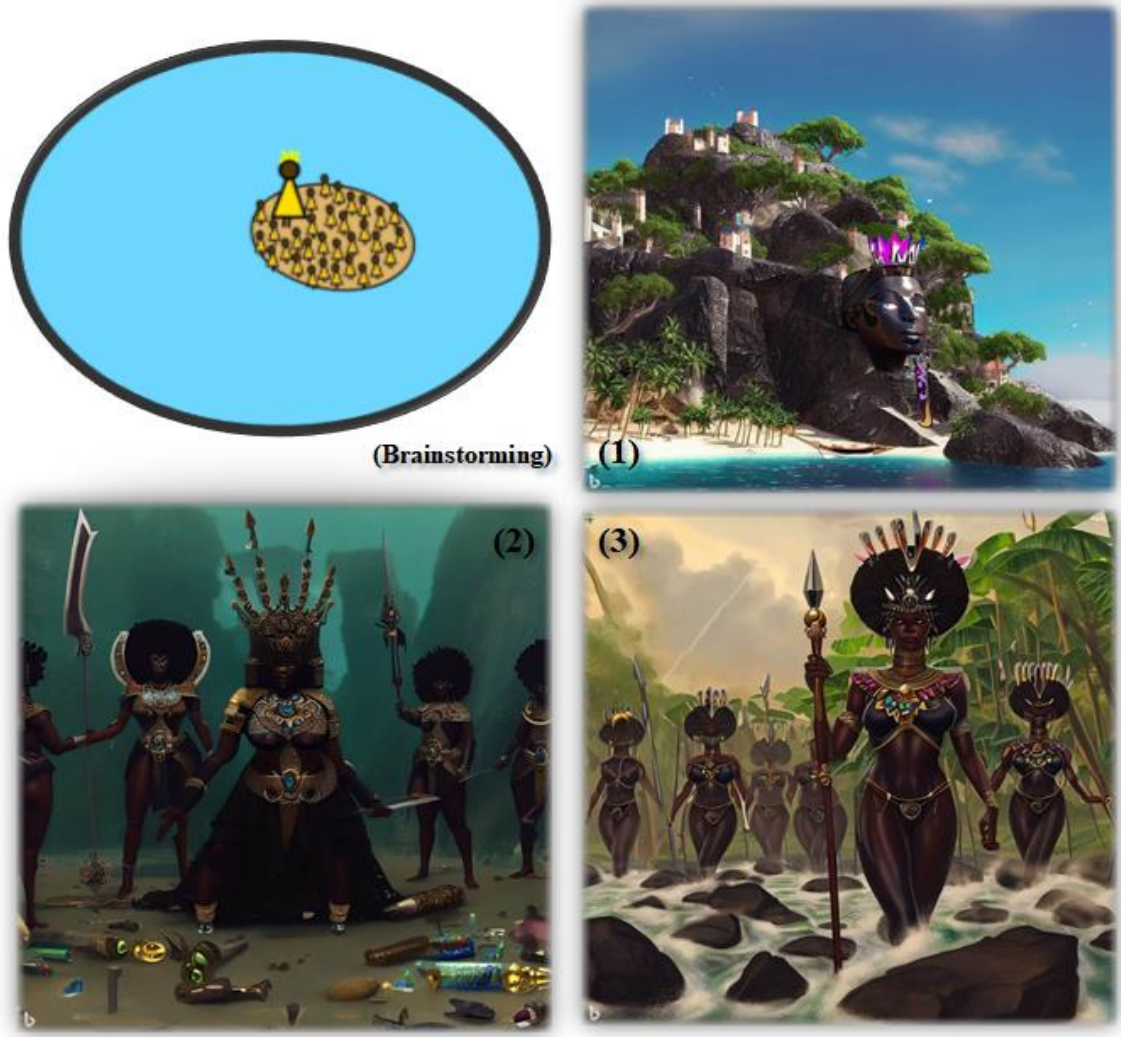

**Figure 5.** Sample images associated with "California"

Another participant chose the word "Justice." After learning that during Roman times, promises and justice were highly valued — to the extent that even a king would execute his own son for breaking the law — he sought to visualize this concept. To depict this, the first prompt requested a drawing comprised of a king, looking solemn, on his way to kill reluctantly his lovely son with his own hands. Although the initial result was satisfactory, the student wanted the image to depict more literally the act of the son being stabbed with a sword, so he modified the prompt to include the action of the king "stabbing his son with a sword." After seeing the second image, which felt too cruel, he requested that the son's figure be "turned around" and "lying face down," leading to the final image (see Figure 6).

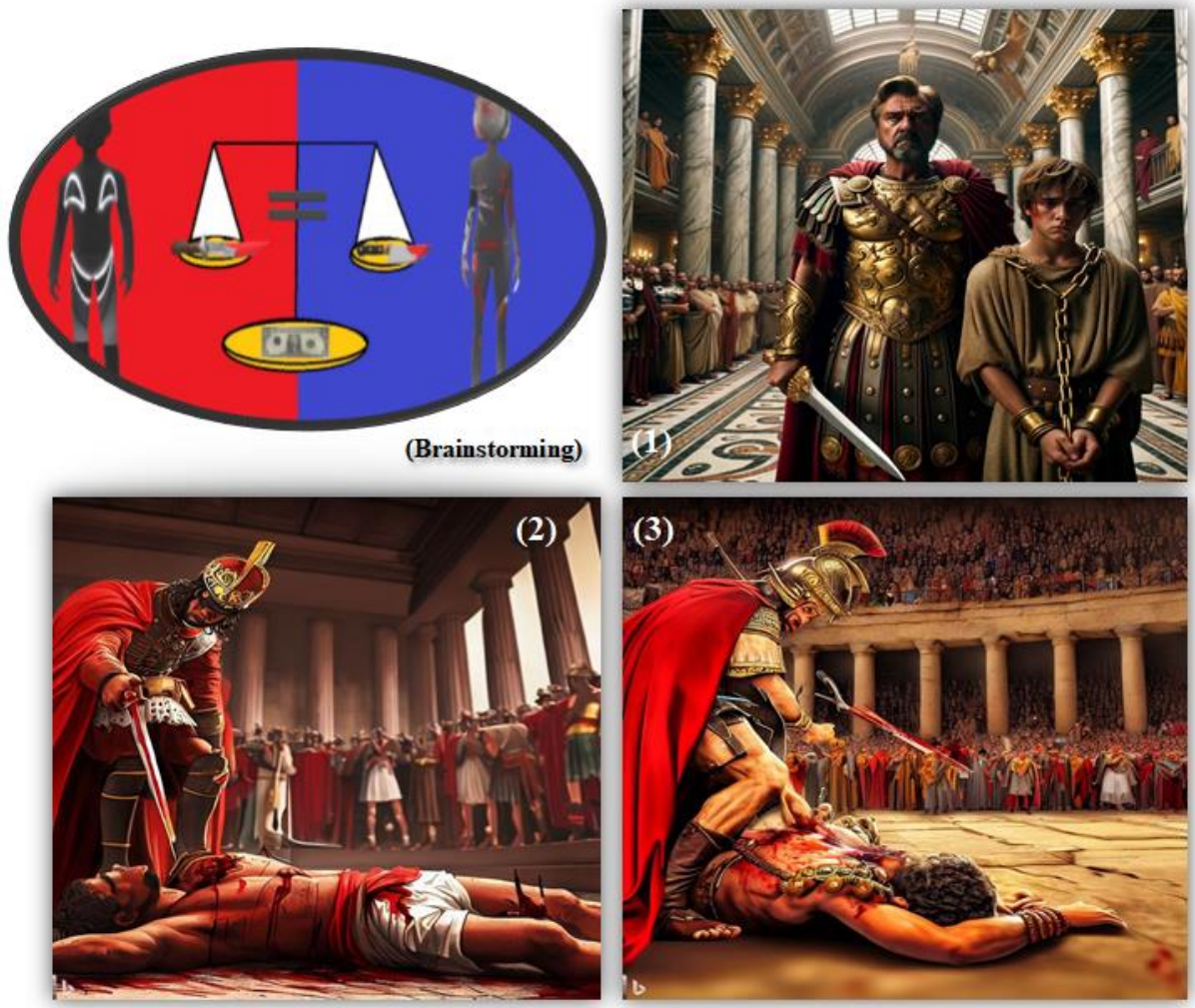

**Figure 6.** Sample images associated with "justice"

Another participant learned that the term "kitsch" originally could mean "trash," but that it also referred to a style in which an outdated fashion, once considered tacky, has been re-designed to create a sense of coolness. To illustrate this, she began by depicting the main character that could represent the word "kitsch." The initial prompt was, "draw a man in anime style wearing a checkered coat, donning a fedora, with a red necktie, and holding a silver cane." However, finding the plain white background too bland and desiring a more realistic depiction, she then added the context, "in a space with a brick-textured wall and wooden flooring," and requested the drawing be in a "realistic style." Although the generated image was satisfactory, it still did not capture the affluent vibe she had initially envisioned, so she added the detail of the character "counting money" to complete the final image (see Figure 7).

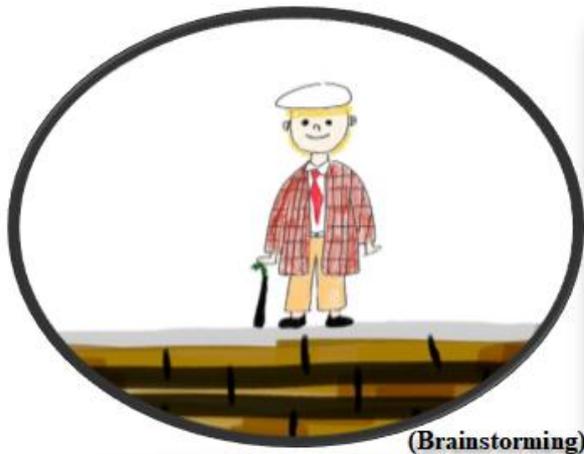
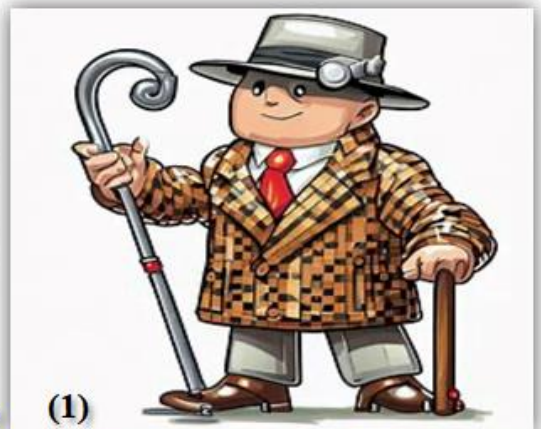
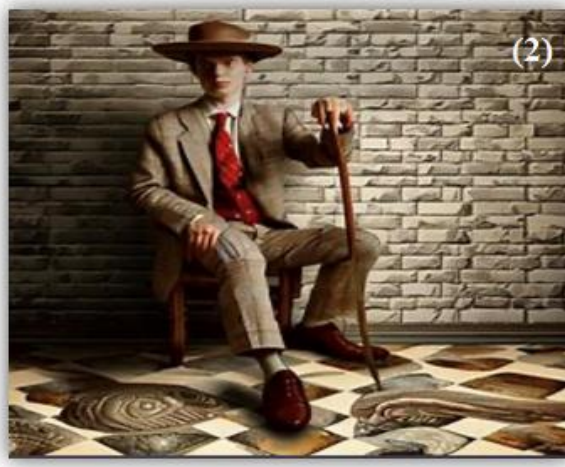
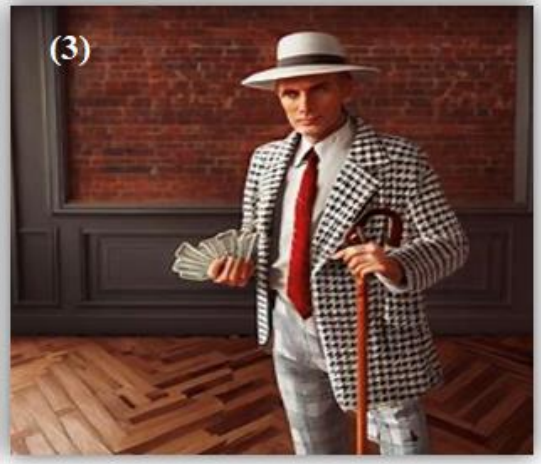

**Figure 7.** Sample images associated with "kitsch"

    The next participant chose the word "barbarian" and depicted a brave warrior figure. While he was pleased with the initial AI-generated image, he noticed the barbarian was not holding a weapon as initially intended by his sketch. He then added the detail "holding a spear." However, in order to create a more masculine appearance, he decided to change the "spear" to an "ax" in the prompt (see Figure 8).

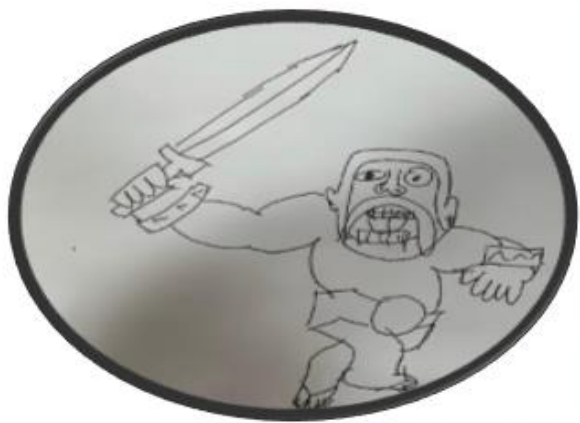
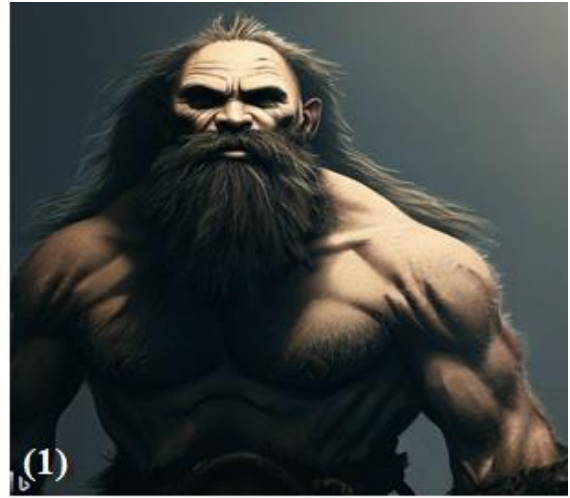
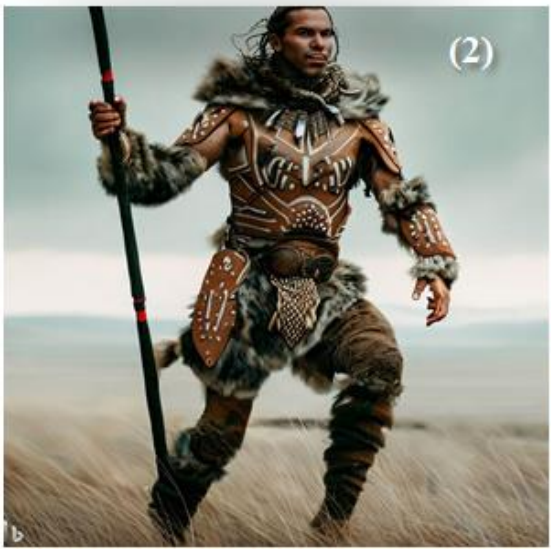
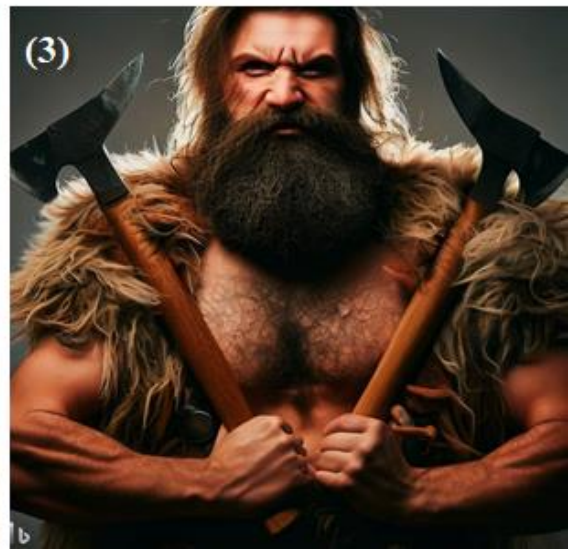

**Figure 8.** Sample images associated with "barbarian"

**4.2 Prompt literacy to activate cognitive vocabulary learning strategies**

The present study further investigated the participants' self-perceived development of cognitive vocabulary learning strategies via the AI-powered vocabulary-image creation project (see Table 2). According to the survey data, as a result of engaging in the target project, the participants judged their uses of the selected set of cognitive vocabulary learning strategies very high overall, with the highest mean ratings being shown in 'hidden meaning' ($M = 6.47$, $SD = 0.82$), 'use in situation' ($M = 6.43$, $SD = 0.86$), 'use in context' ($M = 6.23$, $SD = 1.46$), 'use in sentence' ($M = 6.13$, $SD = 1.14$), and 'imagination' ($M = 6.03$, $SD = 1.23$).

**Table 2.** Participants' self-perceived development of cognitive vocabulary learning strategies via the target project ($n = 30$)

| Strategies | Constructs | M (SD) |
|---|---|---|
| Visual encoding | Q1 Visualization | 6.13(1.04) |
| Use of word-structure | Q2 Hidden meaning | 6.47(0.82) |
| | Q3 Prefix and suffix | 5.73(1.31) |
| | Q4 Word origin | 5.93(1.26) |
| Contextual encoding | Q5 Use in context | 6.23(1.46) |
| | Q6 Use in sentence | 6.13(1.14) |
| | Q7 Use in situation | 6.43(0.86) |
| Activation | Q8 Imagination | 6.03(1.23) |

**4.3 Participants' perception of the concept and value of prompt literacy**

In the context of prompt literacy, the significance of this vocabulary-image creation project extends beyond the students' mere creation focus. Instead, it includes prompt literacy development, which involves interpreting, analyzing, and subsequently modifying the outcomes generated by generative AI programs. This iterative process continues until students attain their desired results. The participants acknowledged the educational value of prompt literacy during this iterative journey, leading to the identification of three major themes. First, they expressed their enjoyment of the process, which allowed them to enter and adjust information through prompt literacy to manifest the envisioned outcome.

> *Initially, I had wanted the generative AI to produce an image exactly like the sketch I had drawn by hand. However, unless I had crafted a very detailed and specific prompt for that drawing, the resulting image didn't match my imagination. In trying to get the image I wanted, I revised the prompt multiple times and learned how to construct a prompt effectively.* (Participant #7)

> *To create the desired image, I wrote and erased countless sentences, repeating the process multiple times. I kept trying and experimenting, producing a variety of images based on my prompts. When an image finally matched the original image I had in mind, I felt proud and accomplished.* (Participant #2)

> *In order to get the images that I wanted, I had to keep modifying the prompt with the help of ChatGPT. Fine-tuning the prompt to finally produce the image I desired was quite satisfying.* (Participant #14)

> *When I fed the image I had imagined and sketched in my mind into the generative AI, I initially provided a somewhat vague representation of my mental picture. While the result resembled my imagination to some extent, it wasn't perfect. However, when I provided very detailed and specific prompts, the output was almost identical to what I had envisioned, which was quite astonishing.* (Participant #3)

Second, another common theme that emerged from the provided reflection data is the importance of prompt literacy in various aspects of life. Some participants emphasized how the ability to formulate clear and precise requests or responses, akin to using the right prompt, is essential for achieving desired outcomes. They recognize that prompt literacy extends beyond the context of generative AI and facilitates effective communication, problem-solving, and success in academic and professional settings. The theme emerged around the idea that prompt literacy, as a broader skill, is crucial for personal growth and career development in an era marked by rapid technological advancement. This is seen in the following excerpts:

> *Through prompt literacy, I was reminded of the profound importance of words. It became evident that to attain the desired outcomes, I needed to express my intentions clearly and precisely. This experience reinforced the significance of organizing my thoughts and effectively conveying my desires.* (Participant #16)

> *Just as I must input the correct prompt to achieve the desired result, I recognized that in everyday life, the capacity to accurately articulate my thoughts and opinions through language is pivotal in attaining my objectives.* (Participant #21)

> *Prompt literacy refers to the understanding of a specific topic or given situation and the ability to respond appropriately to it. I felt that there's a need for prompt literacy skills in various real-life situations. For instance, in assignments or exams, understanding the given question and providing the appropriate answer is essential to obtain satisfactory scores. Also, adopting the right attitude and response in somewhat tense and formal situations, such as interviews or presentations, plays a crucial role in achieving successful outcomes. Thus, I have learned how to adeptly adapt to circumstances, including using correct prompts. I am convinced that prompt literacy significantly influences not only my personal growth but also my career development.* (Participant #9)

Third, another common theme identified in the responses is the nature of prompt literacy in the context of interacting with generative AI technology. In fact, several participants mentioned in their reflection papers the importance of prompt literacy from a new perspective of human-AI interaction and collaboration. The following excerpts speak to this well:

> *I learned that prompt literacy involves the capacity to provide precise commands to generative AI, directing it to locate the information you require. Additionally, it encompasses the skill of collaborating effectively with generative AI to achieve your goals or tasks efficiently. This combination of command and collaboration is crucial in harnessing the full potential of AI technology.* (Participant #5)

> *Prompt literacy involves the capacity to issue precise commands to generative AI, directing it to retrieve specific information, as well as the proficiency to collaborate effectively with the generative AI. These competencies are of utmost importance, as they enable individuals to provide clear directives for accessing particular information and harness the potential of generative AI to its fullest extent.* (Participant #7)

> *I think prompt literacy necessitates a humanistic sensibility, understanding how to craft the appropriate prompts, and the ability to express these prompts effectively in one's own language to achieve the desired outcomes.* (Participant #10)

> *I believe that the significance of this skill lies in its relevance to our future interactions*

*with AI. Prompt literacy is crucial for ensuring seamless communication in this evolving landscape.* (Participant #15)

Overall, these responses underscore the holistic nature of prompt literacy, encompassing technical proficiency, effective communication, and adaptability in an increasingly AI-driven world.

## 5. Discussion

By implementing the project using generative AI to produce relevant images for English words, the participants developed and demonstrated their prompt literacy skills, as evidenced in the Results section. It is important to note that they developed two common strategies when undertaking the current project: (1) crafting and adjusting prompts to depict precisely the context in which the target word is used, and (2) elaborating on the actions of the characters/objects that relate to the story concerning the target word. The survey data findings also indicate that L2 learners developed a series of cognitive learning strategies, which they can engage in their vocabulary learning. In addition, upon observing L2 learners revisiting and refining their prompts consistently, it is inferred that prompt literacy can aid them in visualizing a word's historical origin and sociocultural meanings. This finding is consistent with well-established cognitive theories like dual coding theory (Paivio, 1986, 2010) and the CTML (Mayer, 2009), both of which highlight the importance of delivering information through both verbal and visual channels to improve understanding and retention. Prompt literacy, in particular, can make this advantage more accessible and comprehensive. In addition, the process of fine-tuning prompts can assist learners in their incidental language learning by facilitating task-induced engagement (Laufer & Hulstijn, 2001). When learners actively engage in the task of refining prompts, they become more deeply immersed in the language learning process. This heightened engagement not only enhances their grasp of the specific vocabulary but also contributes to a more immersive and effective language learning experience overall.

The AI-powered vocabulary-image creation project underscored the importance of prompt literacy, which encompasses not only instructing students in prompt creation and modification but also analyzing the outcomes generated by AI systems. Through the iterative process of refining prompts to align with learners' creative visions, students gained valuable insights into the educational benefits of prompt literacy, its broader impact on personal growth and communication, and the emergence of a collaborative approach to human-AI interaction. This aligns with numerous studies supporting the idea that generative AI in education fosters a culture of co-creation between humans and AI (Jeon & Lee, 2023; Oppenlaender, 2023; Vartiainen & Tedre, 2023). In conjunction with prompt literacy, generative AI's functionality goes beyond simply providing answers; it actively co-designs the learning experience and outcomes. This framework not only guides the development and implementation of generative AI tools in education but also emphasizes that the goal is not to replace human-centric educational practices but to enhance them.

In light of this study's results, we offer the following pedagogical implications. First, we suggest that the AI-powered vocabulary-image creation project, as introduced in the present study, is an ideal activity to achieve two pedagogical goals simultaneously: the development of one's prompt literacy and one's vocabulary learning strategies. Therefore, instructors in the English teaching domain can use this project (or one slightly adapted to the target learners' linguistic needs and cognitive maturity) for this purpose. For learners with relatively lower levels of English proficiency, instructors are encouraged to provide sample prompts and associated AI-generated images, and to demonstrate carefully the entire process of AI-powered image generation. Second, once target students have become familiar with the image creation process, instructors can extend this project to creating a book of stories around a target word, which may lead to the further development and sophistication of students' prompt literacy. Finally, instructors themselves can use AI-powered image generation as one of their pedagogical resources to explain a target word's multiple meanings and its use in varying contexts (Nation, 2001), in addition to providing a verbal explanation.

The present study is not without limitations, two of which are addressed below. First, given the participants' major (English Language and Literature), it is likely that the development of prompt literacy skills demonstrated by the participants is partially, if not largely, owing to their English proficiency level. Thus, it remains unknown whether a similar outcome could be achieved with those of a lower English proficiency level. Second, there were occasions when Bing Image Creator and Dall-E generated different images when the same (or slightly modified) prompts were entered, causing some difficulty for participants in terms of generating an intended image. It remains unexplored to what extent such a technical difficulty negatively affected the participants' engagement in the target project. We believe it would be a worthwhile endeavor to investigate whether the ability to overcome such a difficulty could be an important component of prompt literacy.

## 6. Conclusion

The evolution of literacy reflects societal shifts and technological progress. Traditional literacy, once confined to reading and writing skills, has transformed into digital literacy with the advent of computers and the internet. This expanded concept encompasses information literacy, which emphasizes data retrieval and media literacy, which focuses on using digital media tools (Gilster, 1997). As AI has advanced, a new era of AI literacy has emerged, highlighting the importance of comprehending AI's mechanisms and of proficiently creating and applying related software (Ng et al., 2021). Currently, as we continue to move more deeply into the era of generative AI, "prompt literacy" is emerging as a crucial skillset. While AI literacy highlights one's comprehension of how AI works and coding proficiency, prompt literacy centers on interacting with generative AI to produce specific, desired outcomes. In terms of output, while digital literacy enables content creation based on sourced information, prompt literacy fosters the creation of new results through language input.

Prompt literacy conceptualized and operationalized in this study ensures that AI tools respect and cater to the diverse ways in which students represent and engage with knowledge. In the context of discovery learning (van Joolingen, 1998), prompt literacy creates an environment in which students can explore and learn on their own. While revising prompts and interacting with AI, learners can engage in exploration, experimentation, and discovery. They can collaborate with AI to generate new content, address problems, or engage in creative projects. In this way, prompt literacy provides a comprehensive framework for understanding how AI can be used to support, guide, and challenge students, while also respecting their unique learning journeys. Given such an emergence of common themes in this study, we propose that prompt literacy deserves more attention in the field of AI literacy, as a distinct and emergent ability to develop.

**References**

Al-Seghayer, K. (2001). The effect of multimedia annotation modes on L2 vocabulary acquisition: A comparative study. *Language Learning & Technology*, *5*(1), 202–232. https://doi.org/10125/25117

Andretta, S. (Ed.). (2007). *Change and challenge: Information literacy for the 21st century*. Auslib Press.

Bawden, D. (2001). Information and digital literacies: A review of the concepts. *Journal of Documentation*, *57*(2), 218–259. https://doi.org/10.1108/EUM0000000007083

Bozkurt, A. (2023). Generative artificial intelligence (AI) powered conversational educational agents: The inevitable paradigm shift. *Asian Journal of Distance Education*, *18*(1), 198–204. https://doi.org/10.5281/zenodo.7716416

Bruner, J. S. (1983). *Child's talk: Learning to use language.* Norton.

Burniske, R. W. (2007). *Literacy in the digital age*. Corwin Press.

Cho, S. Y. (2020). *Story of English words*. Kimyoungsa.

Cohen, A. D. (1987). The use of verbal and imagery mnemonics in second-language vocabulary learning. *Studies in Second Language Acquisition*, *9*(1), 43–61. https://doi.org/10.1017/S0272263100006501

Dale, E. (1969). *Audio-visual methods in teaching*. Dryden Press.

Debes, J. L. (1969). The loom of visual literacy. *Audiovisual Instruction*, *14*(8), 25–27.

Dewey, J. (1938). *Experience and education.* Macmillan.

Gersten, R., & Baker, S. (2000). What we know about effective instructional practices for English-language learners. *Exceptional Children*, *66*(4), 454–470. https://doi.org/10.1177/001440290006600402

Gilster, P. (1997). *Digital literacy*. Wiley Computer Publications.

Giray, L. (2023). Prompt engineering with ChatGPT: A guide for academic writers. *Annals of Biomedical Engineering*, 1–5. https://doi.org/10.1007/s10439-023-03272-4

Gu, Y., & Johnson, R. K. (1996). Vocabulary learning strategies and language learning outcomes. *Language Learning*, *46*(4), 643–679. https://doi.org/10.1111/j.1467-1770.1996.tb01355.x

Hsieh, H. F., & Shannon, S. E. (2005). Three approaches to qualitative content analysis. *Qualitative Health Research*, *15*(9), 1277–1288. https://doi.org/10.1177%2F1049732305276687

Hull, G., & Moje, E. B. (2012, April). What is the development of literacy the development of? Paper presented at the Understanding Language Conference, Stanford, CA. Retrieved from http://ell.stanford.edu/papers

Hwang, Y. (2023). The emergence of generative AI and PROMPT literacy: Focusing on the use of ChatGPT and DALL-E for English education. *Journal of the Korea English Education Society, 22*(2), 263-288.

Hwang, Y., Shin, D., & Lee, H. (2023). Students' perception on immersive learning through 2D and 3D metaverse platforms. *Educational Technology Research and Development*, 1-22. https://doi.org/10.1007/s11423-023-10238-9

Jeon, J., & Lee, S. (2023). Large language models in education: A focus on the complementary relationship between human teachers and ChatGPT. *Education and Information Technologies*. https://doi.org/10.1007/s10639-023-11834-1

Jewitt, C. (2008). Multimodality and Literacy in school classrooms. *Review of Research in Education*, *32*(1)*,* 241–267. https://doi.org/10.3102/0091732X07310586

Jewitt, C. (2012). *Technology, literacy, learning: A multimodal approach*. Routledge.

Johns, A. M. (1997). *Text, role, and context: Developing academic literacies.* Cambridge University Press.

Kong, S. C., Cheung, W. M. Y., & Zhang, G. (2021). Evaluation of an artificial intelligence literacy course for university students with diverse study backgrounds. *Computers and Education: Artificial Intelligence*, *2*, 100026. https://doi.org/10.1016/j.caeai.2021.100026

Kost, C. R., Foss, P., & Lenzini Jr, J. J. (1999). Textual and pictorial glosses: Effectiveness on incidental vocabulary growth when reading in a foreign language. *Foreign Language Annals*, *32*(1), 89–97. https://doi.org/10.1111/j.1944-9720.1999.tb02378.x

Laufer, B., & Hulstijn, J. (2001). Incidental vocabulary acquisition in a second language: The construct of task-induced involvement. *Applied linguistics*, *22*(1), 1–26. https://doi.org/10.1093/applin/22.1.1

Lee, S. M. (2023). Second language learning through an emergent narrative in a narrative-rich customizable metaverse platform. *IEEE Transactions on Learning Technologies*. https://doi.org/10.1109/TLT.2023.3267563

Liang, J. C., Wang, J. Y., & Chou, S. H. (2020, September). A designed platform for programming courses. In 2020 9th International Congress on Advanced Applied Informatics (IIAI-AAI) (pp. 290–293). IEEE.

Lo, L. S. (2023). The CLEAR path: A framework for enhancing information literacy through prompt engineering. *The Journal of Academic Librarianship*, *49*(4), 102720. https://doi.org/10.1016/j.acalib.2023.102720

Mayer, R. E. (2009). *Multimedia learning* (2nd ed). Cambridge University Press.

Nation, I. S. P. (2001). *Learning vocabulary in another language*. Cambridge University Press.

Ng, D. T. K., Leung, J. K. L., Chu, S. K. W., & Qiao, M. S. (2021). Conceptualizing AI literacy: An exploratory review. *Computers and Education: Artificial Intelligence*, *2*, 100041. https://doi.org/10.1016/j.caeai.2021.100041

Oppenlaender, J. (2023). Prompt engineering for text-based generative art. *arXiv preprint arXiv*:2204.13988.

Paivio, A. (1986). *Mental representations: A dual coding approach*. Oxford University Press

Paivio, A. (2010). Dual coding theory and the mental lexicon. *Mental Lexicon*, *5*(2), 205–230. https://doi.org/10.1075/ml.5.2.04pai

Paivio, A., & Desrochers, A. (1981). Mnemonic techniques in second-language learning. *Journal of Educational Psychology*, *73*(6), 780–795. https://doi.org/10.1037/0022-0663.73.6.780

Rousseau, M. K., Tam, B. K. Y., & Ramnarain, R. (1993). Increasing reading proficiency of language-minority students with speech and language impairments. *Education and Treatment of Children*, *16*(3), 254–271. Retrieved from https://www.jstor.org/stable/42899316

Saldaña, J. (2013). *The coding manual for qualitative researchers*. Sage Publications.

Steinbauer, G., Kandlhofer, M., Chklovski, T., Heintz, F., & Koenig, S. (2021). A differentiated discussion about AI education K-12. *KI-Künstliche Intelligenz*, *35*, 131–137. https://doi.org/10.1007/s13218-021-00724-8

Su, J., Ng, D. T. K., & Chu, S. K. W. (2023). Artificial intelligence (AI) literacy in early childhood education: The challenges and opportunities. *Computers and Education: Artificial Intelligence*, *4*, 100124. https://doi.org/10.1016/j.caeai.2023.100124

The New London Group. (2000). A pedagogy of multiliteracies: Designing social futures. In B. Cope & M. Kalantzis (Eds.), *Multiliteracies: Literacy learning and the design of social futures* (pp. 9–37). London: Routledge.

van Joolingen, W. (1998). Cognitive tools for discovery learning. *International Journal of Artificial Intelligence in Education*, *10*, 385–397.

Vartiainen, H., & Tedre, M. (2023). Using artificial intelligence in craft education: Crafting with text-to-image generative models. *Digital Creativity*, *34*(1), 1–21. https://doi.org/10.1080/14626268.2023.2174557

Wu, J. G., Zhang, D., & Lee, S. M. (2023). Into the brave new metaverse: Envisaging future language teaching and learning. *IEEE Transactions on Learning Technologies*, 1–11. https://doi.org/10.1109/TLT.2023.3259470

Yanguas, I. (2009). Multimedia glosses and their effect on L2 text comprehension and vocabulary learning. *Language Learning & Technology*, *13*(2), 48–67. https://doi.org/10125/44180